\documentclass[showpacs,amsmath,amssymb,aps,floatfix,pra,twocolumn]{revtex4-1}

\usepackage{graphicx}
\usepackage[colorlinks=true,citecolor=blue,linkcolor=magenta]{hyperref}

\newcommand{\ssm}{\scriptscriptstyle\rm}
\renewcommand{\phi}{\varphi}
\renewcommand{\theta}{\vartheta}
\newcommand{\pdag}{\phantom{\dag}}

\font\tenmib=cmmib10
\font\eightmib=cmmib10 scaled 800
\font\sixmib=cmmib10 scaled 667
\newfam\mibfam

\def\mib{\fam\mibfam\tenmib}
\textfont\mibfam=\tenmib
\scriptfont\mibfam=\eightmib
\scriptscriptfont\mibfam=\sixmib

\def\nd{^{\vphantom{\dagger}}}
\def\ns{^{\vphantom{*}}}
\def\yd{^\dagger}
\def\frac#1#2{{\textstyle{#1 \over #2}}}

\def\ie{{\it i.e.\/}}

\def\half{\frac{1}{2}}

\def\beq{\begin{equation}}
\def\eeq{\end{equation}}
\def \be{\begin{equation}}
\def \ee{\end{equation}}
\def \bea{\begin{eqnarray}}
\def \eea{\end{eqnarray}}
\def\half{\mbox{$1\over2$}}

\def\cC{{\cal C}}
\def\cH{{\cal H}}

\def\cA{{\cal A}}

\def\cF{{\cal F}}
\def\cT{{\cal T}}

\def\bR{{\mib R}}

\def\bA{{\mib A}}

\def\bv{{\mib v}}

\def\bx{{\mib x}}
\def\br{{\mib r}}
\def\by{{\mib y}}

\def\sxy{\sigma_{xy}}

\def\tV{{\textsf V}}

\def\bT{\boldsymbol{\Theta}}

\begin{document}
\title{Topological Transitions for Lattice Bosons in a Magnetic Field}
\author{Sebastian D. Huber$^{1}$ and Netanel H. Lindner$^{2,3}$}
\affiliation{$^1$ Department of Condensed Matter Physics, The Weizmann
Institute of Science, Rehovot, 76100, Israel}
\affiliation{$^2$ Institute for Quantum Information, California Institute of Technology, Pasadena, CA 91125, USA.}
\affiliation{$^3$ Department of Physics, California Institute of Technology, Pasadena, CA 91125, USA.}

\date{\today}

\begin{abstract}
The Hall response provides an important characterization of strongly
correlated phases of matter. We study the Hall conductivity of interacting
bosons on a lattice subjected to a magnetic field. We show that for any
density or interaction strength, the Hall conductivity is characterized by a
single integer. We find that the phase diagram is intersected by topological
transitions between different integer values.  These transitions lead
to surprising effects, including sign reversal of the Hall conductivity
and extensive regions in the phase diagram where it acquires a negative
sign. This implies that flux flow is reversed in these regions - vortices
there flow upstream. Our finding have immediate applications to a wide
range of phenomena in condensed matter physics, which are effectively
described in terms of lattice bosons.
\end{abstract}

\maketitle

The Hall response is a key theoretical and experimental tool for
characterizing emergent charge carriers \cite{Ziman} in strongly
correlated systems, ranging from high temperature superconductors
\cite{Hagen90,LeBoeuf07,LeBoeuf11} to the quantum Hall effect
\cite{Wen95}. In this paper, we study the Hall conductivity of strongly
correlated bosons on a lattice. We find that the entire phase diagram
of such systems can be characterized using a single integer $p$, and
inevitably contains topological transitions between different $p$ -
values. These observations allow us to calculate the Hall conductivity
throughout the whole phase diagram, and we show that they lead to surprising
consequences, such as sign reversals of the Hall conductivity. The model
we study describes a wide range of systems in condensed matter physics,
to which our results have immediate implications.  Examples are cold atoms
on optical lattices \cite{Jaksch98, Jaksch05}, Josephson junction arrays
\cite{Fazio01}, granular superconductors \cite{Simanek79,Doniach81},
and perhaps even high temperature superconductors such as the underdoped
cuprates \cite{Uemura89,Micnas95,Mihlin09}.

In the absence of disorder and at weak magnetic fields the Hall conductivity
of bosonic systems is dominated by the flow of superfluid vortices. For
a continuum (Galilean invariant) superfluid, vortex flow gives a Hall
conductivity which is proportional to the ratio of the particle density and
the applied magnetic field. We find that on the lattice, vortex dynamics
is strongly modified. As a result the Hall conductivity is characterized,
in addition to the particle density, by the integer $p$. We show how
emergent particle-hole symmetry points in the ground-state phase diagram
necessarily lead to a non-trivial behavior of this integer and we discuss
the topological transitions between different $p$-sectors. As we shall show,
these transitions are attributed to degeneracies in the many body spectrum,
which serve as sources for the Berry curvature.

\begin{figure}[t]
\begin{center}
\includegraphics{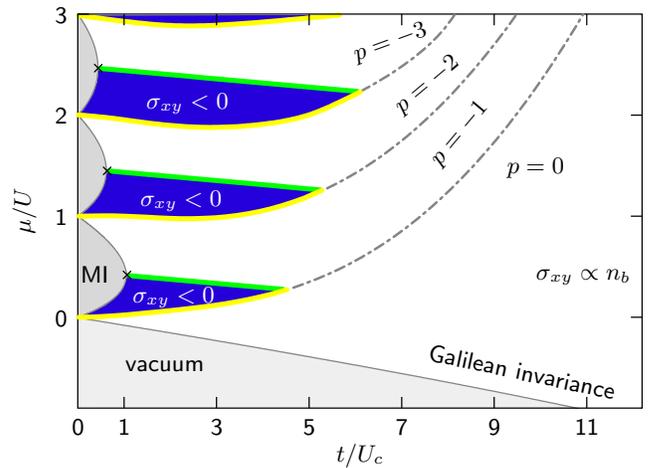}
\end{center}
\caption{
{\bf Topological transitions in the Bose-Hubbard phase diagram.} The
Galilean invariant regime denotes the region where $\sxy$ is proportional
to the particle density $n_b$ divided by the magnetic field strength
$B$. Mott insulator lobes are indicated in gray.  The yellow and green
lines exhibit an emergent particle hole symmetry, where $\sxy=0$. They
are divided into two types: (i) Lines emanating from the tip of the Mott
lobe at integer boson filling where $\sxy$ has a smooth zero crossing
(green). (ii) Transition lines (yellow), through which $B\sxy$ exhibits
integer jumps. The latter continue into the phase diagram, with $\sxy>0$,
as indicated by the dashed lines. The blue region corresponds to regions
where the Hall conductivity is negative.
}
\label{fig:sketch}
\end{figure}

Specifically, we focus on the conventional Bose-Hubbard model \cite{Fisher89}
in two dimensions. We restrict our study to a dissipation-less system,
at zero temperature and without disorder. Within the phase diagram of
this model we find large parameter regions corresponding to a negative
Hall conductivity, $\sigma_{xy}<0$, and reversed vortex motion where
vortices flow upstream, cf.~Fig~\ref{fig:sketch}.  We discuss methods to
directly test these results in cold atom systems where the neutral atoms
are subjected to synthetic magnetic fields introduced through rotation or
phase imprinting \cite{Lin09, Cooper11}.

\section{Hall conductivity and vortex motion}
\label{sec:Hall}

\begin{figure}[t]
\begin{center}
\includegraphics{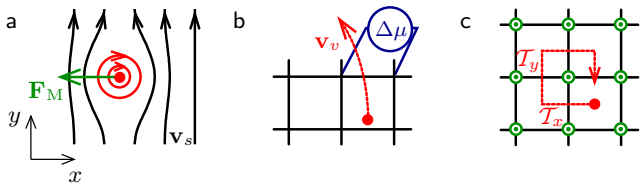}
\end{center}
\caption{
{\bf Forces acting on a vortex.} (a) The classical Magnus force due to
the interaction of the velocity field of the vortex and the external flow
${\bf v}_{s}$ acts perpendicular to ${\bf v}_{s}$.  (b) Vortex motion leads
to a change in the momentum of the system due to its phase singularity,
which is perpendicular to its velocity ${\bf v}_{v}$.  (c) Moving a vortex
around a lattice site yields a Berry phase of $2\pi\alpha=2\pi(n_{b}+p)$.
}
\label{fig:magnus}
\end{figure}

We begin by giving a description of vortex dynamics in bosonic
systems. A vortex moving with respect to a current experiences a force
arising from the interaction of the velocity field of the vortex with
the one of external current. This hydrodynamical force is called the
Magnus force, and it acts perpendicularly to the current, as depicted in
Fig.~\ref{fig:magnus}(a). Similarly, a superfluid vortex (of unit vorticity)
in two dimensions experiences a force
\begin{equation}
{\bf F}_{\ssm M} = - 2\pi \hbar n_{s} {\bf v}_{s} \times
\hat{\bf e}_{z}.
\label{eq:magnus}
\end{equation}
where $n_s$ is the number density of superfluid bosons, and $\bv_s$ is
their velocity. The unit vector $\hat{\bf e}_{z}$ is a normal to the plane.

The force ${\bf F}_{\ssm M}$ in Eq.~(\ref{eq:magnus}) arises from the
dynamical phase (time integral of the energy) in a Lagrangian describing the
superfluid. Such a Lagrangian necessarily contains a term corresponding to
the Berry phase picked up by the vortex motion. The Berry phase acquired
by a vortex moving around a loop of area $S$ is given by $2\pi\alpha S$,
where $\alpha$ is a proportionality factor which depends on the microscopic
details of the Hamiltonian. Therefore, an equation of motion for the
vortex leading to dissipation-less flow is linear in the vortex velocity
and given by \cite{Fisher91}
\begin{equation}
{\bf F}_{\ssm M} + 2\pi \hbar \alpha\, {\bf v}_{v} \times \hat
{\bf e}_{z}=0.
\label{eqn:pl}
\end{equation}

Equation~(\ref{eqn:pl}) can also be understood from the
perspective of momentum balance. A moving vortex imprints a phase
discontinuity on the superfluid wave function. The Josephson relation
$\Delta\mu=\hbar\partial_{t}\Delta\phi$ connects the resulting chemical
potential to the time derivative of the relative phase difference,
cf. Fig.~\ref{fig:magnus}(b).  The chemical potential drop will be balanced
by a flow of particles, which results in momentum transfer from the particles
to the moving vortex, perpendicular to the vortex velocity ${\bf v}_{v}$
and proportional to its magnitude. The proportionality factor $\alpha$
relates the change in the system's momentum to the vortex velocity.

The Hall conductivity can be related to the drift velocity of a vortex. From
(\ref{eq:magnus}) and (\ref{eqn:pl}) we get
\be
\label{eqn:vv}
{\bf v}_{v} = \frac{n_{s}}{\alpha} \bf{v}_{s}.
\ee
In a system with a low density of vortices we can neglect the effects of
vortex-vortex interactions. Considering strictly dissipation-less flow,
we obtain from the Josephson relation a semi-classical expression for the
Hall conductivity
\begin{equation}
\sigma_{xy}=\frac{q n_{s}v_{s}}{\Delta\mu}
=\frac{q^2 n_{s}v_{s}}{ 2\pi \hbar  n_{v}v_{v}}
=\frac{q^2}{h}\frac{\alpha}{n_{v}}.
\label{eq: hall alpha}
\end{equation}
where $n_{v}$ is the density of vortices and $q$ the boson charge.

In systems with Galilean invariance, in a reference frame moving
at the vortex velocity, there should be no forces acting on the
vortex. This requires $\bv_v=\bv_s$ and therefore sets $\alpha=n_{s}$
\cite{Haldane85,Ao93}. This relation is modified in the presence of a
lattice, as we discuss now.

We consider the standard model for interacting bosons on a
lattice~\cite{Fisher89}
\begin{eqnarray}
\label{eqn:BHH}
\cH=&-&t\sum_{\langle \br ,\br'\rangle}
\Bigl[
 b_{\br}^{\dag}b_{\br'}^{\pdag}e^{iA_{\br \br'}}
 + b_{\br}^{\dag}b_{\br'}^{\pdag}e^{-iA_{\br \br'}}
\Bigr]\nonumber\\
&+&\frac{U}{2}\sum_{\br}
b_{\br}^{\dag}b_{\br}^{\pdag}(b_{\br}^{\dag}b_{\br}^{\pdag}-1)
-\mu\sum_{\br} b_{\br}^{\dag} b_{\br}^{\pdag},
\end{eqnarray}
where $b_{\br}^{\dag}$ creates a boson on site $\br$, $t$
is the hopping amplitude, $U$ the on-site repulsion; $A_{\br
\br'}=q\int_{\br}^{\br'}\bf{A}\cdot {\bf dx}$ is the phase factor due
to applied gauge field ${\bf A}$. We work in units where $\hbar=c=1$,
and likewise we set the lattice constant $a=1$.

We first note that vortices live on the center of the plaquettes of
the lattice. We explicitly construct operators $\mathcal T_{x}$ and
$\cT_y$  which translate a vortex by one lattice constant in the $x$
and $y$ directions. We show that they obey the commutation relation (see
supplementary materials for a full derivation)
\begin{equation}
\label{eqn:around}
\mathcal T_{x}\mathcal T_{y}
= \mathcal T_{y} \mathcal T_{x} e^{2\pi i \hat{N_b}/N}.
\end{equation}
Here, $\hat{N}_b $ is the  total boson number operator, and $N$ is the
number of sites. We denote the particle filling by $n_b=N_b/N$. From
Eq.~(\ref{eqn:around}), we see that the Berry phase acquired by moving a
vortex around a dual lattice plaquette is
\begin{equation}
\label{eqn:aa}
2\pi\alpha = 2\pi(n_b + p)\qquad \mbox{with} \qquad p \in
\mathbb{Z}.
\end{equation}
The integer $p$ arises from the $2\pi$ ambiguity in Eq.~(\ref{eqn:around}).

Equation~(\ref{eqn:aa}) can also be understood in terms of momentum balance
\cite{Oshikawa00}. Following Paramekanti and Vishwanath \cite{Paramekanti04},
we note that Eq.~(\ref{eqn:around}) implies that when a vortex is transported
by $\Delta y$ sites along $\hat {\bf e}_{y}$, the momentum of the system
changes by $\Delta P_{x}=2\pi n_b \Delta y$.  At the same time we can
integrate Eq.~(\ref{eqn:pl}) to obtain $\Delta P_{x}= 2\pi\alpha\Delta
y$. Combining the two results and taking into account that momentum is
only conserved up to a reciprocal lattice vector $2\pi p$ leads us to
Eq.~(\ref{eqn:aa}).

The above results, together Eq.~(\ref{eq: hall alpha}), imply a similar
relation for the Hall conductivity,
\beq
\sxy \,n_v= \frac{q^2}{h} \left(n_b + p\right) \qquad\mbox{with} \qquad p \in \mathbb{Z}.
\label{eq: hall aa}
\eeq
While  Eqs.~(\ref{eq: hall alpha}) and (\ref{eq: hall aa}) are a
semi-classical derivation of the Hall conductivity, in Sec.~\ref{sec:away}
we derive an {\em exact} relation between $\alpha$ and the Hall conductivity
for a system containing {\em one vortex},
\beq
\sxy= \frac{q^2}{h} N\alpha=\frac{q^2}{h}N(n_b + p) \qquad\mbox{with} \qquad p \in \mathbb{Z}.
\label{eq:halloneflux}
\eeq

In the remainder of this paper we investigate the relations
(\ref{eqn:aa}--\ref{eq:halloneflux}) throughout the phase diagram of the
Bose-Hubbard model. In Sec.~\ref{sec: low} we study these relations in the
Gross-Pitaevskii and Mott transition limits. In Sections \ref{sec: hcb},
\ref{sec:away} we study the transition between different $p$-sectors in
the hard core boson limit. We complete the phase diagram using numerical
calculations in Sec.~\ref{sec:evolution}.

\section{Low-energy limits}
\label{sec: low}

We start by discussing low energy limits of Eq.~(\ref{eqn:BHH}) where a
diverging length-scale enables the derivation of a continuum low-energy
theory. In these limits $\alpha$ and $\sxy$ can be deduced directly. We
review the derivation of the low-energy theories for weak ($U\ll t$)
and strong ($U\gg t$) interactions. In both cases we start by rewriting
(\ref{eqn:BHH}) as a coherent state path integral with the following action
for the complex valued field $\psi_{i}$
\begin{multline}
\label{eqn:action}
S=\int\!\! d\tau\, \Bigg\{\sum_{i}\psi_{i}^{*}(\partial_{\tau}-\mu)\psi_{i}-
t\sum_{\langle i,j \rangle}(\psi_{i}^{*}\psi_{j}e^{-i\phi_{ij}}+{\rm c.c.})\\
+\frac{U}{2}\sum_{i} |\psi_{i}|^{2}(|\psi_{i}|^{2}-1)\Bigg\}.
\end{multline}

{\bf The Gross-Pitaevskii limit -- } In the weakly interacting limit,
the Gross-Pitaevskii healing length $\xi_{\ssm GPE}= a \sqrt{t/Un_b}$ is
much larger than the lattice spacing $a$. This enables a straight-forward
gradient expansion of (\ref{eqn:action}). To lowest order in gradients we
obtain the continuum action
\begin{equation}
\label{eqn:gpe}
S= \int\!\!d\tau d{\bf x}\, \Big\{\psi^{*}\partial_{\tau}\psi
+ a^{2}t|(\nabla-i q {\bf A}) \psi|^{2} +\dots\Big\}.
\end{equation}

Using the above expression we can now derive the coefficient
$\alpha$ in the Gross-Pitaevskii limit.  When written in terms of
$\psi=\sqrt{n_b}\exp(i\theta)$, the term $\psi^{*}\partial_{\tau}\psi$ in
Eq.~(\ref{eqn:gpe}) leads to a purely imaginary  ${\mathcal L}_{\tau}=2\pi i
m n_b$, where $m$ is a field that counts the winding of the phase $\theta$
\cite{Fisher91}.  Consider the action of a field configuration associated
with taking a vortex around a closed loop of area $S$. A little reflection
shows that the regions outside the loop do not change the value of $m$ while
those inside contribute unity per particle. Hence, $\mathcal{L}_{\tau}$
gives rise to a Berry phase of $2\pi i n_b S$ \cite{Haldane85}. This
observations fixes
\begin{equation}
\alpha=n_b.
\end{equation}

{\bf Around the Mott insulator --} At strong interactions and integer
filling, $n_b \in {\mathbb N}$, the Hamiltonian (\ref{eqn:BHH}) stabilizes
a localized Mott insulating phase with vanishing superfluid fraction
$\bar\psi_{i}\!\equiv\!\langle \psi_{i} \rangle\!=0$ \cite{Fisher89}. In
the insulating phase all sites are occupied by exactly $n_b$ bosons. Both
the addition or removal of a particle is protected by a finite gap. This
gap closes at the boundary of the Mott lobes in the phase diagram of
Fig.~\ref{fig:sketch}, whereby at the lower (upper) boundary of the Mott
lobe the hole (particle) gap vanishes. Hence, the tip of the Mott lobe
represents a multi-critical point where the particle and hole gap close
simultaneously \cite{Capogrosso-Sansone07}. In the following we focus on
this multi-critical point.

When both the particle and hole gap vanish, an enhanced symmetry in
the low energy sector emerges. Instead of going through the standard
procedure of deriving the low-energy theory from microscopic considerations
\cite{Oosten01,Polkovnikov05} we motivate the effective action via its
symmetry properties. We expect the following particle-hole symmetry (PHS)
to hold $\bar \psi \rightarrow \bar\psi^{*}$ and ${\bf A}\rightarrow -{\bf
A}$ \cite{Dorsey92}.  To leading order in powers of $\bar \psi$ we find
\begin{equation}
\label{eqn:mi}
S= \int\!\!d\tau d{\bf x}\, \Big\{\frac{1}{8tn_b}|\partial_{\tau}\bar \psi|^{2}
 + a^{2}tn_b|(\nabla-i q {\bf A}) \bar\psi|^{2}+\dots\Big\}.
\end{equation}
The gradient expansion leading to an effective continuum theory is
controlled by the diverging correlation length close to the second order
phase transition into the Mott insulating state.

A direct consequence of PHS in the {\em continuum theory} (\ref{eqn:mi})
is $\sigma_{xy}({\bf A})=\sigma_{xy}(-{\bf A})$. Together with the
Onsager relation $\sigma_{xy}({\bf A})=-\sigma_{xy}(-{\bf A})$ we obtain
$\sigma_{xy}=0$. This result can also be understood in terms of vortex
motion.  As opposed to the Gross-Pitaevskii action, the continuum theory
(\ref{eqn:mi}) is real. Hence it does not give rise to any Berry phase
when a vortex is moved around a closed loop and we conclude
\begin{equation}
\alpha = 0.
\end{equation}

Starting from the PHS points, we expect to find lines with $\alpha=0$
in the $\mu/U$--$t/U$ phase diagram of the Bose Hubbard model.  From
Eq.~(\ref{eqn:aa}) on the other hand, we know that at a fixed density,
$\alpha$  can only change by an integer. This leads to the conclusion that
the lines with $\alpha=0$ are bound to lines of integer fillings in the
phase diagram, c.f.  Fig.~\ref{fig:sketch}.

\section{Hard core bosons limit}
\label{sec: hcb}

We now consider the limit $t/U\to 0$ and $\mu/U \to m$, where $m$ is an
integer. In Fig.~\ref{fig:sketch}, these limits lie in-between two Mott
lobes. The two states with $m$ and $m+1$ bosons per site are degenerate
single-site states  of Hamiltonian~(\ref{eqn:BHH}). States with different
fillings are separated by a gap of order $U$, and do not appear in the
low energy theory.

We use a Schrieffer-Wolf transformation \cite{macdonald88} to project
the Hamiltonian (\ref{eqn:BHH}) onto the subspace with only $m$ and $m+1$
bosons per site. The resulting Hamiltonian corresponds to hard-core bosons
(HCB) and can be written using spin-$\half$ operators; $S^z_i+\half$
is the onsite number operator, and $S^+_i$ ($S^-_i$) raises (lowers) the
occupation from $m$ to $m+1$ (and vice versa). At zeroth order in $t/U$,
the HCB Hamiltonian is given by
\beq
\cH_{\ssm HC}^{(0)}=  -(m+1) t  \sum_{\langle\br,\br'\rangle}
\left(  e^{iA_{\br \br'} } S^+_\br S^-_{\br'} + {\rm h.c.} \right)- \mu  \sum_{\br} S^z_{\br} .
\label{xxz}
\eeq

The HCB Hamiltonian~(\ref{xxz}) has an emergent charge conjugation
symmetry. One defines the unitary transformation
\be
C  \equiv  \exp\Big(i\pi  \sum_\br S^x_\br \Big).
\ee
$C$ transforms particles into holes, \ie, $C\yd S^z_\br C
 = -S^z_\br$, and
\be
C^\dagger\, \cH^{(0)}_{\ssm HC}\big(q\bA, \mu\big)\,C  =
\cH^{(0)}_{\ssm HC}\big(-q\bA, -\mu\big).
\label{Part-hole}
\ee
At half filling for the hard core bosons, the Hamiltonian~(\ref{xxz}) is
independent of $\mu$ and hence Eq.~(\ref{Part-hole}) implies invariance
under  ${\bf A}\to -{\bf A}$. Hence, the Onsager relation $\sigma_{xy}({\bf
A})=-\sigma_{xy}(-{\bf A})$ implies that for half integer fillings
($n_b=\half+m$)
\begin{equation}
\label{eqn:alphahc}
\sxy=\alpha=0.
\end{equation}

Note that the situation at the HCB limits and at the tip of the Mott lobes
are qualitatively different. The Hall conductivity at the tip of the Mott
lobe vanishes due to a zero-crossing of $\alpha$ when $p=-n_b$. In the HCB
limit, the integer $p$ jumps exactly at $n_b=m+\half$. In other words:
in the first case it reflects a particle hole-symmetry between $n_b-1$
and $n_b+1$ while in the latter the symmetry connects $n_b-1$ and $n_b$ at
$n_b=m+\half$. The symmetry at the HCB limits has a remarkable consequence
for $\sxy$ in the full phase diagram of the model, as we shall now show.

\section{Away from the hard core boson limit}
\label{sec:away}

We now consider the effect of a finite but small value of $t/U$. Second
order processes in which a virtual excitation with an on-site occupation
of $m-1$ or $m+2$ bosons are created lead to corrections to the
Hamiltonian~(\ref{xxz}) of order $t^2/U$.  Taking into account all the
different processes we obtain up to irrelevant re-normalizations of the
parameters in $\cH_{\ssm HC}$
\begin{equation}
\cH_{\ssm HC}^{(1)}\!=\!\cH^{(0)}_{\ssm HC}-
\epsilon_m\!\!\!\sum_{\langle\langle \br,\br';\br''\rangle\rangle}\!\!\!
\left(e^{iA_{\br \br'}} S_{\br}^+
S_{\br'}^-(S^z_{\br''}+\half)+{\rm h.c.}\right),
\label{eq: hcb1}
\end{equation}
where $\langle\langle \br,\br';\br''\rangle\rangle$ denote sites
$\br$ and $\br'$ which are nearest neighbors of site $\br''$, and
$\epsilon_m=(m+1)(m+2) t^2/U$.

The new terms in  Eq.~(\ref{eq: hcb1}) break the charge conjugation symmetry,
and therefore  the Hall conductivity at exactly half integer filling does
not vanish; below we calculate it in the limit of small $t/U$.

We consider the model Eq.~(\ref{eq: hcb1}) on a torus of size $N=L_{x}L_{y}$,
with $N$ even. The gauge field ${\bf A}$ describes a uniform flux penetrating
the surface of the torus. We take the total flux to be one flux quantum,
which induces one vortex into the system. An important gauge invariant
quantity described by the gauge field are the two Wilson line functions
\cite{Lindner10}
\beq
\Phi_x(y)=\oint dx A_x ,\qquad \Phi_y(x)=\oint dy A_y.
\label{eq: wilsontext}
\eeq
We define $\Theta_x=\Phi_x(y=0)$ and $\Theta_y=\Phi_y(x=0)$.  Changing the
values of $\Theta_x$ and $\Theta_y$ corresponds to threading Aharonov-Bohm
(AB) fluxes through the two holes of the torus \cite{Lindner10}.

The Hall conductivity at zero temperature, for a general many-body
Hamiltonian can be calculated by integrating the Berry curvature
\cite{Avron85}
\beq
\sxy= {q^2\over h}{1 \over 2\pi  }
\int\limits_0^{2\pi}\!\!d\Theta\ns_x
\!\!\int\limits_0^{2\pi}\!\!d\Theta\ns_y\>\cF.
\label{eq: sxy}
\eeq
$\cF$ is given by
\be
\cF=\epsilon^{\mu \nu}\partial_{\Theta_\mu}A_{\nu},\qquad \cA_\nu=i\left\langle{
\Psi_0 (\bT) }\bigg| {\partial \Psi_0(\bT)\over\partial
\Theta\ns_\nu}\right\rangle.
\label{eq: curvature}
\ee
Here $\Psi_0(\bT)$ is the many-body ground state wave function which
depends on the Aharonov-Bohm fluxes through the holes of the torus.

Remarkably, the Hall conductivity in the presence of one vortex can be
calculated analytically at half filling. The key ingredients are degeneracies
in the spectrum that occur for $t/U=0$ \cite{Lindner09,Lindner10} and serve
as point (monopole) sources for the Berry curvature $\cF$ \cite{Simon83}.

To understand the degeneracies, we consider an effective Hamiltonian for the
vortex hopping between dual lattice sites. As shown in Ref.~\cite{Lindner09,
Lindner10}, this is given by
\begin{multline}
\cH_\tV  =     - t\ns_\tV  \sum_{\langle \bR,\bR'\rangle} \left( e^{i
A^{\rm D}_{\bR,\bR'} } \,  b\yd_{\bR} b\nd_{\bR'} +
\mbox{h.c.}\right)
\\+
\sum_{\bR}U({\bf R}-{\bf R}_{\rm V}) \,b\yd_\bR b\nd_\bR,
\label{eq: hb}
\end{multline}
where $b_{\bf R}^{\dag}$ creates a vortex on a dual lattice site, and $t_{\rm
V}\approx t$. The dual gauge field's flux is given by the boson density,
$\nabla \times {\boldsymbol A^{\rm D}}=\Phi^{D} = 2\pi n_b$. The potential
$U(\bR-\bR_{\rm V})$ for the vortex position arises due to the fact that
the Wilson lines (\ref{eq: wilsontext}) break translational symmetry on
the torus. In fact, as shown in \cite{Lindner09,Lindner10} for one flux
quantum penetrating the surface of the torus, all translational symmetries
are absent, and the potential $U(\br-\bR_{\rm V})$ acquires its minimum at a
point $\bR_{\rm V}$ for which the Wilson lines both take on the value $\pi$.
\begin{figure}[t]
\begin{center}
\includegraphics{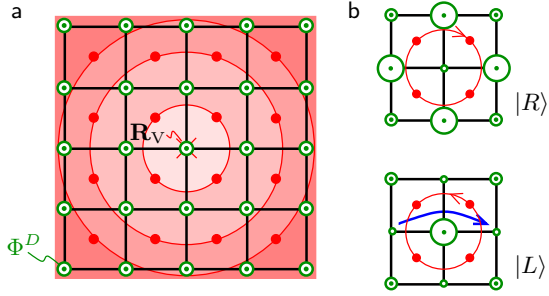}
\caption{
{\bf Vortex Hamiltonian.} (a) The vortex hops between dual lattice sites
(red), where at half filling for the bosons the average dual flux per
plaquette is $\half$ a flux quantum. The circular contours represent  the
equipotential contours for the confining potential for the vortex. When
the Aharonov-Bohm fluxes are tuned to $\Theta_x^0,\Theta_y^0$, the minimum
of the potential is situated on a (direct) lattice site and the vortex
ground state has two degenerate states $|R\rangle$, $|L\rangle$. (b)
The two ground states correspond to two different charge density wave
configurations, with $|R\rangle$ ($|L\rangle$) corresponding to depleted
(excess) charge at $\bR_{\rm V}$. When the particle hole symmetry breaking
terms of Eq.~(\ref{eq: hcb1}) are introduced, hopping terms through $\bR_{\rm
V}$ (depicted by the blue arrow) lower the energy of the state $|L\rangle$
relative to $|R\rangle$.
}
\label{fig: vortex ham}
\end{center}
\end{figure}

If the point $\bf R_{\rm V}$ lies on a  site of the direct lattice,
the eigenstates of $\cH_\tV$ (in a symmetric gauge) can be written
as $\psi(\bR-\bR_{\rm V})=f(\left|\bR-\bR_{\rm V}\right|)e^{i m
\phi(\bR-\bR_{\rm V})}$. Here $\phi\left(\bR-\bR_{\rm V}\right)$
denotes the angle between $\bR-\bR_{\rm V}$ and the $x$-axis, and
$m=0,\pm 1,\pm 2$. At half filling, the average dual flux per plaquette
is $\overline{\Phi}_{D}=\pi$ and the ground state is doubly degenerate
with $m=0, 1$. The two states $|R\rangle$ ($m=0$) and $|L\rangle$ ($m=1$)
represent states with clockwise and counter-clockwise vortex currents,
respectively, as depicted in Fig.~\ref{fig: vortex ham}(b). Note that this
two-fold degeneracy occurs for $N$ distinct values of $\bT$.

To calculate $\sxy$ we need to analyze the spectrum around the $N$
degeneracy points. Around these points, the Hamiltonian restricted to
the $|R\rangle$ and $|L\rangle$ basis is of the form $H_{\rm V}={\bf
h}\cdot {\boldsymbol \sigma}$ . To find ${\bf h}$, we first notice that if
$\bT^0=\left(\Theta_x^0,\Theta_y^0\right)$ is a degeneracy point, tuning away
from it by $\bT=\bT^0+\Delta\bT$, moves $\bR_{\rm V}$ as \cite{Lindner10}
\beq
\bR_{\rm
V}=\bR^0_{\rm V}+\Delta\bR_{\rm V},\qquad\Delta R_{\rm V}^\alpha=
-\frac{1}{2\pi}\epsilon^{\alpha\beta} L_\alpha \Delta\Theta_\beta,
\label{eq:vortexmove}
\eeq
where $\alpha,\beta=x,y$ and indices are not summed. Thus, tuning away from
$\bT^0$ breaks the degeneracy between the two ground states $|R\rangle$
and $|L\rangle$, as it shifts the minimum of the potential $U(\bR-\bR_{\rm
V} )$.  Second, the terms $\propto \epsilon_{m}$ in (\ref{eq: hcb1}) lift
the degeneracy even at $\bT=\bT^0$; the assisted hopping terms through
$\bR_{\rm V}$ (blue arrow in Fig.~\ref{fig: vortex ham}) favor $|L\rangle$
over $|R\rangle$.

Together, these two effects give rise to the following low energy Hamiltonian
for each degeneracy point (see supplementary materials for details),
\beq
\cH_\tV= \tilde{U} \left(-\Delta\Theta_y \sigma_x
+\Delta\Theta_x \sigma_y\right) + \tilde{\epsilon} \sigma_z.
\label{eq: hamdeg}
\eeq
where the energy scales appearing above are $\tilde{U}\propto\frac{\partial
U}{\partial{\bR}}\frac{L_{\alpha}}{2\pi}$, and $\tilde{\epsilon}
\propto\epsilon_m$.

We use Eqs.~(\ref{eq: sxy}-\ref{eq: hamdeg}) to calculate the Hall
conductivity for {\em one vortex}.  Consider the Hamiltonian $\cH_{\ssm
HC}^{(1)}$ of Eq.~(\ref{eq: hcb1}) where we let the parameter $\epsilon_m$
take on both negative and positive values. For $0<\epsilon_m/t \ll 1$,
the many body ground state $\Psi_+(\bT,\epsilon_{m})$ is non-degenerate,
and likewise $\Psi_-(\bT,\epsilon_{m})$ for $-1\ll\epsilon_{m}/t<0$. For
$\epsilon_{m}=0$ the two states become degenerate at a set of $N$ values
of $\bT$ space. The Berry connection of the ground state manifold at
$\epsilon_m=0$ is given by
\be
\cA_\mu=i\sum_{i=+,-}
\left\langle{
\Psi_i (\bT) }\bigg| {\partial \Psi_i(\bT)\over\partial
\Theta\ns_\mu}\right\rangle
\label{eq: curvature sum}
\ee
and must satisfy $\int d^2\bT\,\cF=0$ due to particle hole symmetry at
$\epsilon_m=0$. As a result,
\beq
\sxy(\epsilon_{m}>0)+\sxy(\epsilon_{m}<0)=0
\label{eq: sum}
\eeq
\begin{figure}[t]
\begin{center}
\includegraphics{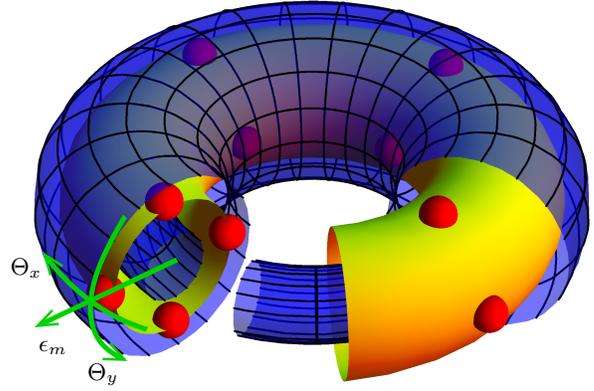}
\caption{
{\bf Berry monopoles. } The parameter space of $\Theta_x$,$\Theta_y$,
and $\epsilon_m$. The yellow surface denotes the particle-hole symmetric
manifold $\epsilon_m=0$, on which the red dots denote the degeneracy
points for $N$ values of $(\Theta_x,\Theta_y)$. The blue surfaces denote
$\epsilon_{m}<0$ and $\epsilon_m>0$. The integral of the curvature $\cF$
on these surfaces counts the number of sources in the three dimensional
parameter space which they enclose.
}
\label{fig: monopoles}
\end{center}
\end{figure}
Next, we consider the space of $\Theta_x,\Theta_y, \epsilon_m$, which has
the topology of a thick torus, as depicted in Fig.~\ref{fig: monopoles}. We
are interested in the integral of the Berry curvature $\cF$ on the surfaces
with $\epsilon_m>0$ and $\epsilon_m<0$, which yield $\sxy(\epsilon_m>0)$
and $\sxy(\epsilon_m<0)$ respectively. At the same time, an analog of Gauss'
law for $\cF$ implies that the integral of  $\cF$ on these surfaces counts
the number of sources for $\cF$ \cite{Berry84,Simon83}. These are just the
degeneracy points discussed above, which are all described by Eq.~(\ref{eq:
hamdeg}) and therefore correspond to sources with charge $+\half$. This
leads to
\beq
\sxy(\epsilon_{m}>0)-\sxy(\epsilon_{m}<0)=N.
\label{eq: diff}
\eeq
Combining Eqs.~(\ref{eq: sum}) and (\ref{eq: diff}) gives
\beq
\sxy(0<t/U\ll 1)=\frac{N}{2}.
\eeq

Before concluding this section, we note that Eq.~(\ref{eq:vortexmove})
leads to an exact relation between $\alpha$ and the Hall conductivity of
one vortex. From Eq.~(\ref{eq:vortexmove}), the Berry phase for moving a
vortex around a plaquette is given by
\beq
2\pi\alpha=\oint_\cC d\Theta_\mu\, \cA_\mu =
\int_{S(\cC)} d^2\bT \, \cF
\eeq
where the contour $\cC$ defines a square of size $2\pi/L_x \times 2\pi/
L_y$ in flux space, and $S(\cC)$ is the surface it bounds.  Therefore,
for one vortex,
\beq
\alpha N=\frac{1}{2\pi}\int\limits_0^{2\pi}\!\!d\Theta\ns_x
\!\!\int\limits_0^{2\pi}\!\!d\Theta\ns_y\, \cF  =
\sxy.
\label{eq:equivalphasxy}
\eeq

Consider again the phase diagram of the Bose Hubbard model. From the above
discussion, we conclude that the transitions lines between two integer
values for $p$ emanate from the HCB points, and move to higher densities with
increasing $t/U$. These lines all correspond to changes of the integer $p$
by unity. The PHS lines emanating from the neighboring Mott lobe tips {\em
terminate} at the transition lines, cf.  Fig.~\ref{fig:sketch}. Together,
they define regions with negative $\alpha$ and Hall conductivity.

\section{Evolution of the transition lines}
\label{sec:evolution}

We numerically calculate the Chern number (\ref{eq: sxy}) for one vortex,
to obtain the behavior of the integer $p$ in the full parameter regime of
the Bose Hubbard model. We use a Lanczos algorithm \cite{Albuquerque07}
to find the ground-state wave function $\Psi_{0}(\bT)$ for different AB
fluxes. Using a standard procedure \cite{Fukui05} to numerically integrate
the Berry curvature (\ref{eq: curvature}) we obtain the Hall conductivity
for different values of $t/U$ and $n_b$.

\begin{figure}[t]
\begin{center}
\includegraphics{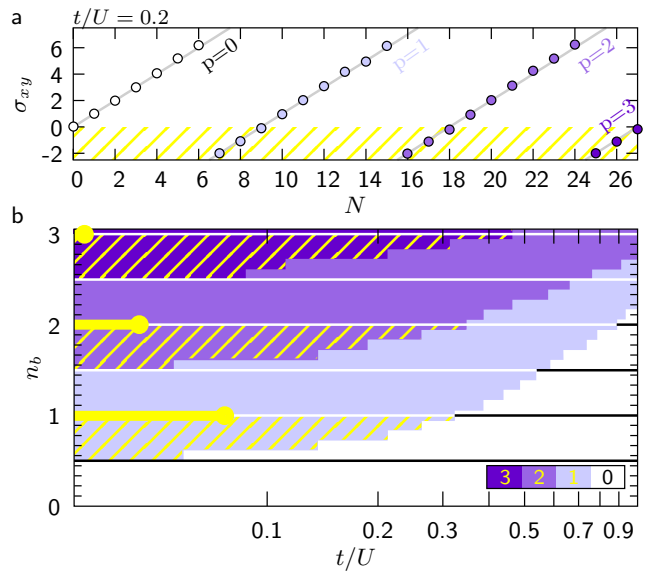}
\end{center}
\caption{
{\bf Numerical results.} (a) The Chern number calculated numerically on
a $3\times 3$ system at $t/U=0.2$ for various particle numbers $N$. The
different branches are described by different integers $p$. The regions
where the resulting Hall conductivity is negative is hatched in yellow. (b)
Phase diagram obtained from cuts of the form of panel (a). Different
colors indicate the different integers $p$. The lines where the integer
$p$ changes by one are the topological transitions. The Mott insulators
at integer filling are indicated by the yellow bars.
}
\label{fig:pdf}
\end{figure}

In Fig.~\ref{fig:pdf} we show the results obtained for a $3\times 3$ cluster,
cf. supplementary materials. We indicate which integer $p$ describes the
Hall conductivity in panel (b). Panel (a) shows a trace of $\sxy$ for
different particle numbers at $t/U=0.2$. In both panels the region in the
phase diagram where $\sxy<0$ are marked with yellow hatches. As expected
from the calculation at half-filling, the transition lines between two
integer value of $p$ move to higher densities for $t/U>0$. Remarkably,
the transition lines intersect the integer density line at increasing
values of $t/U$ for higher densities. As a direct consequence the area of
negative Hall conductivity {\em increases} for higher densities. This is
in contrast to the {\em decreasing} extent of the Mott insulating phases
indicated by the yellow bars in Fig.~\ref{fig:pdf}(b). This surprising
behavior of the Bose-Hubbard model is explained below.

In order to see at which values of $t/U$ a sign change of $\alpha$ should
be expected at integer fillings, we consider the healing length $\xi_{\ssm
GPE}$, which sets the size of a vortex. For $\xi_{\ssm GPE}=a\sqrt{t/Un_b}\ll
a$ the size of a vortex is much smaller then the lattice spacing $a$
\cite{Huber08}, and the Bose Hubbard model maps onto the quantum rotor
model \cite{Simanek80, Doniach81} which has an emergent PHS at integer
filling \cite{Fisher91,Sonin97}. We have seen  that PHS implies $\alpha=0$.
The dependence of $\xi_{\ssm GPE}$ on the {\em mean-field} interaction $Un_b$
therefore explains the growing extent of the negative Hall conductivity. This
has to be contrasted to the {\em bosonic enhancement factors} in the
hopping terms $\propto \sqrt{n_b}$ which facilitate the melting of the
Mott insulator and lead to smaller Mott lobes at increasing densities.

Finally, in Fig.~\ref{fig:sketch} we present the numerical results as a
function of $t/U$ and $\mu/U$. To translate from the results at fixed
density $n_b$ to a fixed chemical potential $\mu$ we use a standard
mean-field approach \cite{Huber07}.

\section{Experimental verification}
\label{sec:exp}

Our results have a direct experimental signature in terms of the vortex
flow velocity in a moving system of lattice bosons. If $\sigma_{xy}$ is
positive (negative) the vortices move with (against) the superfluid flow,
cf. Eq.~(\ref{eqn:vv}).

In a cold-atoms setup, the direction and speed of the vortex flow can be
measured with in-situ imaging techniques \cite{Bakr09}. However, in the
strongly interaction regime the vortex core is smaller then a lattice
spacing $a$. Hence, to make the vortex visible in the density profile,
the system parameters have to be ramped into the weakly interaction regime
before imaging.

The sign change of the Hall conductivity can also be measured studying
collective modes in a trap. When the atom cloud is displaced from the minimum
of the harmonic trap the atoms start to oscillate in the trap \cite{Jin96}. A
non-vanishing $\sigma_{xy}$ induces a transverse force on this dipole mode
leading to a rotation of the axis of oscillation. Depending on the sign
of the Hall conductivity the rotation is clock or counter-clockwise.

\section{Discussion and outlook}
\label{sec:discussion}

In this paper we focused on vortex dynamics for the Bose-Hubbard model. We
mapped the sectors corresponding to different integer $p$ which characterizes
the Hall conductivity and vortex motion throughout the phase diagram. We
found that close to the Mott insulating phases the sign of $\sigma_{xy}$
is reversed and vortices flow against the applied current.

Our results are obtained neglecting vortex-vortex interactions or
disorder. We note, however, that there are an infinite number of particle
hole symmetric points in the zero temperature phase diagram: at the tip of
every Mott lobe, and in between two adjacent Mott lobes. We saw that the
latter necessarily slice the full phase diagram into an infinite number
of different $p$-sectors. This underlying structure cannot be removed
by the inclusion of vortex-vortex interactions or disorder. However,
the transition lines are expected to change their exact location and to
be smoothed out by these effects, as well as by finite temperature.

Incidentally, reversal of the Hall conductivity have been repeatedly measured
in many strongly correlated electronic materials, including high temperature
superconductors, e.g. in Refs.~\cite{Hagen90,LeBoeuf07,LeBoeuf11}. These
experiments are beyond the direct applicability of our model. An extension
to treat these materials is an interesting future direction. As discussed
above, a clean verification of our predictions is possible in systems of
cold atoms.\\

\noindent \textbf{Acknowledgement}

\noindent We thank Assa Auerbach, Ehud Altman, Joseph Avron, Hans-Peter
B\"uchler, Olexi Motrunich, and Ady Stern for fruitful discussions. Special
thanks to Daniel Podolsky for his enlightening comments. NHL acknowledges
support by the Gordon and Betty Moore Foundation through Caltech's
Center for the Physics of Information, National Science Foundation Grant
No. PHY-0803371, and the Israel Rothschild foundation. SDH acknowledges
support by the Swiss Society of Friends of the Weizmann Institute of Science.
This research was supported in part by the National Science Foundation
under Grant No. PHY05-51164.

\appendix

\section{Vortex translation operators}
\label{app:aa}

We derive the commutation relations for the vortex translation operators
\beq
\cT_x \cT_y=\cT_y \cT_x \exp( 2\pi i \hat N_b/N)
\label{eq:result}
\eeq
Our derivation of equations~(\ref{eq:result}) holds for any interaction
strength $U$  in the Bose Hubbard model (5). We consider the Hamiltonian of
Eq.~(5)  on a torus with $N=L_xL_y$ sites. The gauge field $A_{\br \br'}$
describes one flux quantum piercing the surface of the torus uniformly,
whereby the flux per plaquette is given by
\beq
B=\frac{2\pi}{L_xL_y}.
\eeq
An important gauge invariant quantity described by the gauge field ${\bf
A}$ are the Wilson loop functions
\beq
\Phi_x(y)=\oint dx A_x ,\qquad \Phi_y(x)=\oint dy A_y.
\label{eq: wilson}
\eeq
\begin{figure}[!t]
\begin{center}
\includegraphics{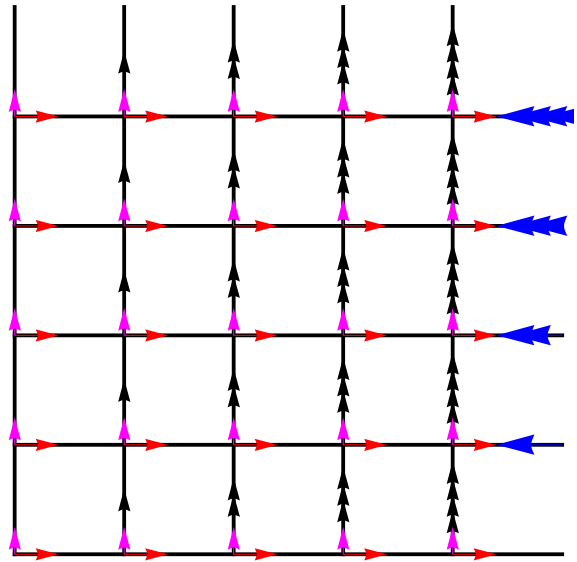}
\caption{
The gauge choice Eq.~(\ref{eq: gauge}). Arrows on links represent the
values for $A_{\br,\br'}$: Black arrows represent a value of $B$, blue
arrows represent $B L_x$, red represent $\Theta_x/L_x$, and magenta
$\Theta_y/L_y$. Open links stand for periodic boundary conditions.
}
\label{fig: gauge}
\end{center}
\end{figure}

We choose a continuous parametrization of the gauge field, which yields
a continuous family of Wilson line function.  Our gauge choice is given by
 \bea
A^x_{\br,\br+\hat{\bx}} &=&y B L_x \delta_{x,L_x-1}+
\frac{\Theta_y}{L_x} \nonumber\\
 A^y_{\br,\br+\hat{\by}}&=& x B L_x +\frac{\Theta_x}{L_y}  .
\label{eq: gauge}
\eea
where $x=0,...,L_x-1$, and $y=0,...,L_{y}-1$. Our  gauge choice is shown
in Fig.~\ref{fig: gauge}. The two Wilson loop functions are given by
\bea
\Phi\ns_y(x) &=&  \phantom{-}  x B L_y +\Theta\ns_y, \\
\Phi\ns_x(y) &=&  - y B L_x +\Theta\ns_x.
\label{eq: wilson loops}
\eea
Note that the parameters $\Theta_x,\Theta_y \in \left[0,2\pi\right]$ define
a \textit{continuous} family of Hamiltonians, which are \textit{inequivalent}
under gauge transformation.

Let $t_x$ and $t_y$ be lattice translation operators, \textit{i.e.}
\beq
 t_x^{\dag} b_{\br} t_x  = b_{\br+\hat{\bf x}}, \qquad t_y^{\dag} b_{\br} t_y  =
 b_{\br+\hat{\bf y}}
\eeq
Translating the Hamiltonian by $t_x$ and $t_y$ leads to
\bea
t_\alpha^\dag \cH[\bA] t_\alpha &=&  \cH[\tilde{\bA}^{(\alpha)}],  \nonumber\\
\tilde{A}^{(\alpha)}_{\br,\br'} &=& A_{t_\alpha(\br),t_\alpha(\br')}.
\eea
The gauge invariant content of the new gauge fields $\tilde{\bA}^{(\alpha)}$
is the same flux per plaquette $B$ as for $\bA$, however the Wilson line
functions of Eq.~(\ref{eq: wilson}) are shifted by one lattice constant,
\bea
t_x : \Phi_y(x) \to \tilde{\Phi}_y(x)&=&\Phi_y(x+1) \nonumber \\
t_y : \Phi_x(y) \to \tilde{\Phi}_x(y)&=&\Phi_x(y+1).
\eea
Note that the values of the Wilson lines cannot be changed by a gauge
transformation. Therefore if we conjugate the Hamiltonian by $t_x$
or $t_y$ we cannot make a gauge transformation back to the original
Hamiltonian. However, we can find gauge transformations $U_x$ and $U_y$
that yield the following relations:
\beq
U_\alpha^{\dag}t_\alpha^{\dag}\cH\left[\bA(\bT)\right]t_\alpha
U_\alpha^{\dag}=\cH\left[\bA\left(\bT-\Delta
\bT^{(\alpha)}\right)\right],
\label{eq: tx}
\eeq
where $\bT=(\Theta_x,\Theta_y)$, and
\beq
(\Delta \Theta^{(\alpha)})_\beta=\epsilon^{\alpha\beta}B
L_{\beta}.
\eeq
where $\epsilon^{\alpha\beta}$ is an antisymmetric tensor with $\epsilon^{x
y}=1$.

We parametrize the unitaries $U_x$ and $U_y$ as
\beq
\nonumber
U_x=\exp(i 2\pi \sum_\br \chi^x_\br \, b^{\dag}_\br b_\br),\quad U_y=\exp(i 2\pi
\sum_\br \chi^y_\br\, b^{\dag}_\br b_\br),
\eeq
where the functions $\chi^\alpha_\br$ are given by
\beq
\chi^\alpha_\br =  \int^{\br} d\br' \cdot
\left(\bA(\bT-\Delta \bT^{(\alpha)})-\tilde{\bA}^{(\alpha)}\right).
\eeq

We now calculate the commutation relation between $\cT_x$ and $\cT_y$. An
explicit formula for $\chi^\alpha_\br$ using the gauge choice Eq.~(\ref{eq:
gauge}) reads
\bea
\chi^x(\br) &=&  -B L_x y \delta_{x,0}, \nonumber\\
\chi^y(\br) &=&  \phantom{-}B x.
\label{eq: chi}
 \eea
Multiplying these operators we get
\bea
\cT_y \cT_x &=& t_y t_x
\exp\left(i\sum_\br\left(\chi^x_\br+\chi^y_{\br-\bx}\right) \,
b^{\dag}_\br b_\br\right),\nonumber\\
\cT_x \cT_y &=& t_x t_y
\exp\left(i\sum_\br\left(\chi^x_{\br-\by}+\chi^y_{\br}\right)\,
b^{\dag}_\br b_\br\right)\nonumber\\
&=&\cT_y \cT_x \exp(i\Upsilon).
\label{eq: comm}
\eea
In the above equation, we have used $\left[t_x,t_y\right]=0$. The factor
$\exp(i\Upsilon)$ is given by
\beq
\Upsilon=\sum_\br \omega_\br \, b^{\dag}_\br b_\br,
\eeq
with
\beq
\omega_\br=\chi^x_\br-\chi^x_{\br-\by}+\chi^y_{\br-\bx}-\chi^y_{\br}.
\eeq
Using Eq.~(\ref{eq: chi}) we get
\beq
\omega_\br=-B,
\eeq
and substituting this into Eq.~(\ref{eq: comm}) we arrive at our final result
\beq
\cT_x \cT_y=\cT_y \cT_x \exp( 2\pi i \hat N_b/N).
\label{eq: around appendix}
\eeq
Here, $\hat N_b$ is the boson number operator and $N$ are the number
of sites.

Note that although explicit gauge choice were made in the derivation of
Eq.~(\ref{eq:result}),the result is gauge invariant: a different gauge
choice in Eq.~(\ref{eq:result}) yield the same result.

To relate $\cT_x$ and $\cT_y$ to the vortex position, we note that the
vortex position can only depend on the values the Wilson lines $\Phi_x(y)$
and $\Phi_y(x)$, as these the only gauge invariant quantities which break
the translational symmetry on the torus.  Therefore, the ground states of
the continuous family of Hamiltonians $\cH[\bA(\bT)]$  corresponds to many
body states $\Psi(\Theta)$ with vortex positions continuously parametrized
by $\bT$ as well. The vortex position in the many body states $\Psi(\bT)$
has quantum fluctuations; the amplitudes however are centered  around a point
$\bR_{\rm V}$ which depends only on $\bT$ \cite{Lindner09,Lindner10}. As
a result, the action of $\cT_x$ and $\cT_y$ shifts the vortex position by
one lattice site in the $x$ and $y$ direction accordingly.

To relate Eq.~(\ref{eq: around appendix}) to $2\pi\alpha$, the Berry phase
acquired by moving a vortex around a dual lattice plaquette, we note that
\bea
\exp(i 2\pi\alpha)&=&\exp(i \oint d\Theta_\mu
\cA_{\mu})\nonumber\\
&=&\langle\Psi(\bT^0)|U_4U_3U_2U_1|\Psi(\bT^0)\rangle,
\label{eq:plaqflux}
\eea
where $\cA_\mu=i\langle \Psi(\bT)|\partial_{\Theta_\mu} \Psi(\bT)\rangle$,
and the line integral is taken around a plaquette in flux space
of size $\left(B L_y,B L_x\right)$. In Eq.~(\ref{eq:plaqflux}),
$U_i=U(\bT^i,\bT^{i+1})$ are adiabatic evolution operators from $\bT^i$
to $\bT^{i+1}$, and
\bea
\bT^1&=&\bT^0+\Delta\bT^{(x)} \nonumber\\
\bT^2&=& \bT^0+\Delta\bT^{(x)}+\Delta\bT^{(y)} \nonumber\\
\bT^3&=& \bT^0+\Delta\bT^{(y)} \nonumber\\
\bT^4&=& \bT^0. \nonumber\\
\eea
Using $U_3=\cT_y U_1^{\dag} \cT_y^\dag$ and $U_4=\cT_x^\dag U_2^{\dag}\cT_x$
and assuming that the ground state $|\Psi(\bT)\rangle$ is non-degenerate,
we find
\bea
\exp(i \oint d\Theta_\mu
\cA_{\mu})&=&\langle\Psi(\bT^0)|
\cT^{\dag}_y\cT^{\dag}_x\cT_y\cT_x|\Psi(\bT^0)\rangle\nonumber\\
&=&\exp(-i 2\pi n_{b})
\label{eq:berryfluxplaq}
\eea
Finally, from Eq.~(\ref{eq:berryfluxplaq}) we see that the Berry flux
through an elementary plaquette in flux space (which has the topology of
a torus) is $2\pi(n_{b}+p)$. All of the $N$ elementary plaquette on the
flux torus are identical. The Hall conductivity is given by the integral of
the Berry curvature on the whole flux torus \cite{Avron85}, and therefore,
in the presence of one vortex,
\beq
\sxy = N(n_{b}+p).
\label{eq: hall result appendix}
\eeq

\section{Vortex hopping Hamiltonian}

\begin{figure}
\includegraphics{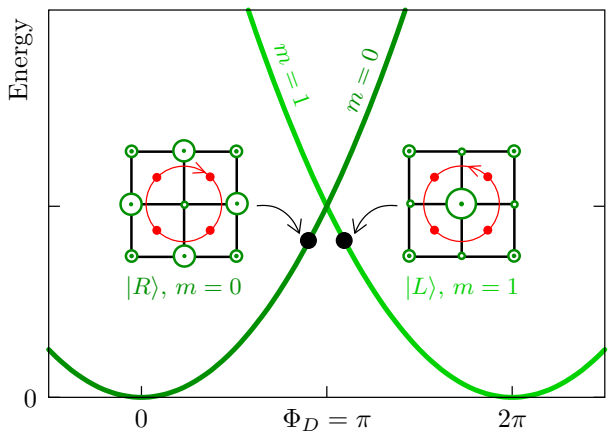}
\caption{
{\bf Vortex Hamiltonian.} The green curves show the energy for a particle
on a ring as a function of the flux through its center.  At slightly
higher (lower) flux than $\Phi_{D}=\pi$ the state $m=1$ ($m=0$) is the
ground state. Two charge density waves centered at $\bR_{\rm V}$ lead to
$\Phi_D=\pi+\delta,\pi-\delta$ and correspond to the states $|L\rangle$,
$|R\rangle$ which have the same energy. Therefore, the state $|L\rangle$
($m=1$) corresponds to a charge density wave with excess density at the
center site.
}
\label{fig:appvort}
\end{figure}

In the following we derive the form of Hamiltonian (25).  Let us start with
the terms arising from a change in the AB fluxes $\bT=\bT^0+\Delta\bT$,
which moves the vortex potential $U(\bR-\bR_{\rm V})$ minimum according to
(24). We can now apply degenerate perturbation theory in the subspace of
$|R\rangle$ and $|L\rangle$. The effect of $\Delta\bT$ only leads to off
diagonal matrix elements between $|R\rangle$ and $|L\rangle$ since both
states have the same $|\psi(\bR-\bR_{\rm V})|$ and the perturbation is
diagonal with respect to $\bR$. The state $|R\rangle+e^{-i\phi}|L\rangle$
has excess weight along $\arctan(x/y)=\phi$, and therefore becomes the
ground state for $\Delta\bR_{\rm V}$ along that direction, i.e.,
\beq
\cH_\tV = \tilde{U} \left(-\Delta\Theta_y \sigma_x
+\Delta\Theta_x \sigma_y\right).
\eeq

We now consider the effect of the particle-hole symmetry breaking terms
of Eq.~(19). We expect the two ground states $|R\rangle$ and $|L\rangle$
to conform to two charge density wave orders centered at $\bR_{\rm
V}$, as the moving vortex exerts a force on the particles due to the
Josephson relation. We note that the two charge density wave orders decay
exponentially with the distance from $\bR_{\rm V}$ (the decay length scale
is the lattice constant) \cite {Lindner09,Lindner10}. To see which of the
states ($|R\rangle$ or $|L\rangle$) has an excess (reduced) density at
$\bR_{\rm V}$, we consider an analogy with  a particle hopping on a ring
around $\bR_{\rm V}$.

In Fig.~\ref{fig:appvort} we show the energy of a particle on a ring
for the two states $|R\rangle$ with $m=0$ and $|L\rangle$ with $m=1$
as a function of the flux $\Phi_D$ through the center of the ring. Note
that the energy of $|R\rangle$ for $\Phi_D=\pi-\delta$ is equal to that
of $|L\rangle$ for $\Phi_D=\pi+\delta$ (at $\Phi_D=\pi$ the two states
are degenerate). Consider now the vortex Hamiltonian of Eq.~(23).
 At half filling for HCBs, we can consider two dual flux
configurations which at $\bR_{\rm V}$ have $\Phi_D(\bR_{\rm
V})=\pi\pm\delta$. Via $\Phi_D(\br)=2\pi\langle S^z_\br+\half \rangle$
they are related to two corresponding charge configurations (the two
configurations are related by charge conjugation). From the analogy to the
particle on a ring, we can infer that these two charge configurations lead to
the vortex ground states $|L\rangle$, $|R\rangle$ respectively. We therefore
conclude that the state $|L\rangle$ ($|R\rangle$) has an excess (reduced)
density at $\bR_{\rm V}$, respectively, cf.  Fig.~\ref{fig:appvort}.

While for the particle-hole symmetric point these charge configurations are
equivalent energetically, the assisted hopping ($t^2/U$) terms in Eq.~(19)
give different energies for the two configurations. Due to the exponential
decay of the charge density wave order \cite {Lindner09,Lindner10}, the
difference between the expectation value of these terms in the states
$|L\rangle$, $|R\rangle$ is also going to decay exponentially with the
distance to $\bR_{\rm V}$.

To account for their effect we estimate the energy change using
mean field HCBs states for $|L\rangle$, $|R\rangle$ of the form
$|\Psi\rangle=\prod_{\br}|\psi_{\br}(\theta_{\br})\rangle$, with
\be
|\psi(\theta_{\br})\rangle = \cos(\theta_{\br}) | \downarrow
\rangle +\sin(\theta_{\br})e^{i\phi_\br} | \uparrow \rangle.
\ee
Here $\phi_\br$ is the phase arising due to the votex. Due to the exponential
decay of the charge density wave order we only consider the $3\times 3$
cluster shown in Fig.~\ref{fig:appvort}.  We evaluate the assisted hopping
(19) in a state where the parameters $\theta_{\br}$ are chosen such
that the center site has $n_{b}=\half\pm\delta$, its nearest neighbors
$n_{b}=\half\mp\delta/4$, and the sites at the corners of the cluster have
$n_{b}=\half$. We find that the energy difference is dominated by assisted
hopping terms which hop over the central site $\bR_{\rm V}$
\begin{multline}
E(|L\rangle)-E(|R\rangle)\approx -\epsilon_{m}\sum_{\br,\br'}\bigg[
\langle L|e^{i A_{\br\br'}}S^+_\br (S^z_{\bR_{\rm V}}+\half)
S^-_{\br'}
|L\rangle\\
-\langle R|e^{i A_{\br\br'}}S^+_\br (S^z_{\bR_{\rm V}}+\half)
S^-_{\br'} |R\rangle \bigg]
\end{multline}
where $\br$ and $\br'$ are nearest neighbors of $\bR_{\rm V}$.  As discussed
above, the state $|L\rangle$ corresponds to higher density at $\bR_{\rm V}$,
therefore the quantity above is negative.

The combined effect of moving the vortex position $\bR_{\rm V}$ away from
a direct lattice site and the PHS symmetry breaking terms of (19) leads
to the low energy effective Hamiltonian near the degeneracy point
\beq
\cH_\tV= \tilde{U} \left(-\Delta\Theta_y \sigma_x
+\Delta\Theta_x \sigma_y\right) + \tilde{\epsilon} \sigma_z,
\eeq
which is given in Eq.~(25).

\section{Exact diagonalization}
\label{app:ED}

We calculate the ground state wave function for different AB fluxes
using the ALPS Lanczos application \cite{Albuquerque07} on a $3\times 3$
cluster. We choose the same gauge choice a depicted in Fig.~\ref{fig:
gauge}. To obtain the phase diagram we truncate the local Hilbert-space
to  include all occupation states up to five particles per site.

To get some insight into finite size effects we also calculate $\sigma_{xy}$
for a $3\times 4$ cluster at filling $n_b=1$. To compare the two cluster
sizes we estimate the Mott transition by considering the gap to the first
excited state. We attribute the transition to  a kink in the gap as a
function of $t/U$. If we rescale the results for the Hall conductivity by
the critical $t/U$ the change from $\sigma_{xy} = 0$ to $\sxy=1$ obtained
with the two clusters fall on top of each other.


\begin{thebibliography}{10}
\expandafter\ifx\csname url\endcsname\relax
  \def\url#1{\texttt{#1}}\fi
\expandafter\ifx\csname urlprefix\endcsname\relax\def\urlprefix{URL }\fi
\providecommand{\bibinfo}[2]{#2}
\providecommand{\eprint}[2][]{\url{#2}}

\bibitem{Ziman}
\bibinfo{author}{Ziman, J.~M.}
\newblock \emph{\bibinfo{title}{Principles of the theory of solids}}
  (\bibinfo{publisher}{Cambridge University Press, London},
  \bibinfo{year}{1972}).

\bibitem{Wen95}
\bibinfo{author}{Wen, X.-G.}
\newblock \bibinfo{title}{Topological orders and edge excitations in fractional
  quantum hall states}.
\newblock \emph{\bibinfo{journal}{Adv. in Phys.}}
  \textbf{\bibinfo{volume}{44}}, \bibinfo{pages}{405} (\bibinfo{year}{1995}).

\bibitem{Hagen90}
\bibinfo{author}{Hagen, S.~J.}, \bibinfo{author}{Lobb, C.~J.},
  \bibinfo{author}{Greene, R.~L.}, \bibinfo{author}{Forrester, M.~G.} \&
  \bibinfo{author}{Kang, J.~H.}
\newblock \bibinfo{title}{Anomalous hall effect in superconductors near their
  critical temperatures}.
\newblock \emph{\bibinfo{journal}{Physical Review B}}
  \textbf{\bibinfo{volume}{41}}, \bibinfo{pages}{11630} (\bibinfo{year}{1990}).

\bibitem{LeBoeuf07}
\bibinfo{author}{LeBoeuf, D.} \emph{et~al.}
\newblock \bibinfo{title}{Electron pockets in the fermi surface of hole-doped
  high-tc superconductors}.
\newblock \emph{\bibinfo{journal}{Nature}} \textbf{\bibinfo{volume}{450}},
  \bibinfo{pages}{533--536} (\bibinfo{year}{2007}).

\bibitem{LeBoeuf11}
\bibinfo{author}{LeBoeuf, D.} \emph{et~al.}
\newblock \bibinfo{title}{Lifshitz critical point in the cuprate superconductor
  yba$_{2}$cu$_{3}$o$_{y}$ from high-field hall effect measurements}.
\newblock \emph{\bibinfo{journal}{Physical Review B}}
  \textbf{\bibinfo{volume}{83}}, \bibinfo{pages}{054506}
  (\bibinfo{year}{2011}).

\bibitem{Jaksch98}
\bibinfo{author}{Jaksch, D.}, \bibinfo{author}{Bruder, C.},
  \bibinfo{author}{Cirac, J.~I.}, \bibinfo{author}{Gardiner, C.~W.} \&
  \bibinfo{author}{Zoller, P.}
\newblock \bibinfo{title}{Cold bosonic atoms in optical lattices}.
\newblock \emph{\bibinfo{journal}{Phys. Rev. Lett.}}
  \textbf{\bibinfo{volume}{81}}, \bibinfo{pages}{3108} (\bibinfo{year}{1998}).

\bibitem{Jaksch05}
\bibinfo{author}{Jaksch, D.} \& \bibinfo{author}{Zoller, P.}
\newblock \bibinfo{title}{The cold atom hubbard toolbox}.
\newblock \emph{\bibinfo{journal}{Annals of Physics}}
  \textbf{\bibinfo{volume}{315}}, \bibinfo{pages}{52--79}
  (\bibinfo{year}{2005}).

\bibitem{Fazio01}
\bibinfo{author}{Fazio, R.} \& \bibinfo{author}{van~der Zant, H.}
\newblock \bibinfo{title}{Quantum phase transitions and vortex dynamics in
  superconducting networks}.
\newblock \emph{\bibinfo{journal}{Phys. Rep.}} \textbf{\bibinfo{volume}{355}},
  \bibinfo{pages}{235} (\bibinfo{year}{2001}).

\bibitem{Simanek79}
\bibinfo{author}{Simanek, E.}
\newblock \bibinfo{title}{Effect of charging energy on transition temperature
  of granular superconductors}.
\newblock \emph{\bibinfo{journal}{Solid State Comm.}}
  \textbf{\bibinfo{volume}{31}} (\bibinfo{year}{1979}).

\bibitem{Doniach81}
\bibinfo{author}{Doniach, S.}
\newblock \bibinfo{title}{Quantum fluctuations in two-dimensional
  superconductors}.
\newblock \emph{\bibinfo{journal}{Phys. Rev. B}} \textbf{\bibinfo{volume}{24}},
  \bibinfo{pages}{5063} (\bibinfo{year}{1981}).

\bibitem{Uemura89}
\bibinfo{author}{Uemura, Y.~J.} \emph{et~al.}
\newblock \bibinfo{title}{Universal correlations between t$_{c}$ and
  n$_{s}$/m$^{*}$ (carrier density over effective mass) in high-t$_{c}$ cuprate
  superconductors}.
\newblock \emph{\bibinfo{journal}{Physical Review Letters}}
  \textbf{\bibinfo{volume}{62}}, \bibinfo{pages}{2317} (\bibinfo{year}{1989}).

\bibitem{Micnas95}
\bibinfo{author}{Micnas, R.}, \bibinfo{author}{Robaszkiewicz, S.} \&
  \bibinfo{author}{Kostyrko, T.}
\newblock \bibinfo{title}{Thermodynamic and electromagnetic properties of
  hard-core charged bosons on a lattice}.
\newblock \emph{\bibinfo{journal}{Physical Review B}}
  \textbf{\bibinfo{volume}{52}}, \bibinfo{pages}{6863} (\bibinfo{year}{1995}).

\bibitem{Mihlin09}
\bibinfo{author}{Mihlin, A.} \& \bibinfo{author}{Auerbach, A.}
\newblock \bibinfo{title}{Temperature dependence of the order parameter of
  cuprate superconductors}.
\newblock \emph{\bibinfo{journal}{Physical Review B}}
  \textbf{\bibinfo{volume}{80}}, \bibinfo{pages}{134521}
  (\bibinfo{year}{2009}).

\bibitem{Fisher89}
\bibinfo{author}{Fisher, M. P.~A.}, \bibinfo{author}{Weichman, P.~B.},
  \bibinfo{author}{Grinstein, G.} \& \bibinfo{author}{Fisher, D.~S.}
\newblock \bibinfo{title}{Boson localization and the superfluid-insulator
  transition}.
\newblock \emph{\bibinfo{journal}{Phys. Rev. B}} \textbf{\bibinfo{volume}{40}},
  \bibinfo{pages}{546--570} (\bibinfo{year}{1989}).

\bibitem{Lin09}
\bibinfo{author}{Lin, Y.~J.}, \bibinfo{author}{Compton, R.~L.},
  \bibinfo{author}{Jimenez-Garcia, K.}, \bibinfo{author}{Porto, J.~V.} \&
  \bibinfo{author}{Spielman, I.~B.}
\newblock \bibinfo{title}{Synthetic magnetic fields for ultracold neutral
  atoms}.
\newblock \emph{\bibinfo{journal}{Nature}} \textbf{\bibinfo{volume}{462}},
  \bibinfo{pages}{628--632} (\bibinfo{year}{2009}).

\bibitem{Cooper11}
\bibinfo{author}{Cooper, N.~R.}
\newblock \bibinfo{title}{Optical flux lattices for ultracold atomic gases}.
\newblock \emph{\bibinfo{journal}{Physical Review Letters}}
  \textbf{\bibinfo{volume}{106}}, \bibinfo{pages}{175301}
  (\bibinfo{year}{2011}).

\bibitem{Fisher91}
\bibinfo{author}{Fisher, M. P.~A.}
\newblock \bibinfo{title}{Hall effect at the magnetic-field-tuned
  superconductor-insulator transition}.
\newblock \emph{\bibinfo{journal}{Physica A}} \textbf{\bibinfo{volume}{177}},
  \bibinfo{pages}{553} (\bibinfo{year}{1991}).

\bibitem{Haldane85}
\bibinfo{author}{Haldane, F. D.~M.} \& \bibinfo{author}{Wu, Y.-S.}
\newblock \bibinfo{title}{Quantum dynamics and statistics of vortices in
  two-dimensional superfluids}.
\newblock \emph{\bibinfo{journal}{Physical Review Letters}}
  \textbf{\bibinfo{volume}{55}}, \bibinfo{pages}{2887} (\bibinfo{year}{1985}).

\bibitem{Ao93}
\bibinfo{author}{Ao, P.} \& \bibinfo{author}{Thouless, D.~J.}
\newblock \bibinfo{title}{Berry's phase and the magnus force for a vortex line
  in a superconductor}.
\newblock \emph{\bibinfo{journal}{Physical Review Letters}}
  \textbf{\bibinfo{volume}{70}}, \bibinfo{pages}{2158} (\bibinfo{year}{1993}).

\bibitem{Oshikawa00}
\bibinfo{author}{Oshikawa, M.}
\newblock \bibinfo{title}{Commensurability, excitation gap, and topology in
  quantum many-particle systems on a periodic lattice}.
\newblock \emph{\bibinfo{journal}{Phys. Rev. Lett.}}
  \textbf{\bibinfo{volume}{84}}, \bibinfo{pages}{1535} (\bibinfo{year}{2000}).

\bibitem{Paramekanti04}
\bibinfo{author}{Paramekanti, A.} \& \bibinfo{author}{Vishwanath, A.}
\newblock \bibinfo{title}{Extending luttinger's theorem to $z_2$ fractionalized
  phases of matter}.
\newblock \emph{\bibinfo{journal}{Phys. Rev. B}} \textbf{\bibinfo{volume}{70}},
  \bibinfo{pages}{245118} (\bibinfo{year}{2004}).

\bibitem{Capogrosso-Sansone07}
\bibinfo{author}{Capogrosso-Sansone, B.}, \bibinfo{author}{Prokov'ef, N.} \&
  \bibinfo{author}{Svistunov, B.}
\newblock \bibinfo{title}{Phase diagram and thermodynamics of the
  three-dimensional bose-hubbard model}.
\newblock \emph{\bibinfo{journal}{Phys. Rev. B}} \textbf{\bibinfo{volume}{75}},
  \bibinfo{pages}{134302} (\bibinfo{year}{2007}).

\bibitem{Oosten01}
\bibinfo{author}{van Oosten, D.}, \bibinfo{author}{van~der Straten, P.} \&
  \bibinfo{author}{Stoof, H. T.~C.}
\newblock \bibinfo{title}{Quantum phases in an optical lattice}.
\newblock \emph{\bibinfo{journal}{Phys. Rev. A}} \textbf{\bibinfo{volume}{63}},
  \bibinfo{pages}{053601} (\bibinfo{year}{2001}).

\bibitem{Polkovnikov05}
\bibinfo{author}{Polkovnikov, A.}, \bibinfo{author}{Altman, E.},
  \bibinfo{author}{Demler, E.}, \bibinfo{author}{Halperin, B.} \&
  \bibinfo{author}{Lukin, M.~D.}
\newblock \bibinfo{title}{Decay of superfluid currents in a moving system of
  strongly interacting boson}.
\newblock \emph{\bibinfo{journal}{Phys. Rev. A}} \textbf{\bibinfo{volume}{71}},
  \bibinfo{pages}{063613} (\bibinfo{year}{2005}).

\bibitem{Dorsey92}
\bibinfo{author}{Dorsey, A.~T.}
\newblock \bibinfo{title}{Vortex motion and the hall effect in type-ii
  superconductors: A time-dependent ginzburg-landau theory approach}.
\newblock \emph{\bibinfo{journal}{Phys. Rev. B}} \textbf{\bibinfo{volume}{46}},
  \bibinfo{pages}{8376} (\bibinfo{year}{1992}).

\bibitem{macdonald88}
\bibinfo{author}{Macdonald, A.~H.}, \bibinfo{author}{Girvin, S.~M.} \&
  \bibinfo{author}{Yoshioka, D.}
\newblock \bibinfo{title}{$t/u$ expansion for the hubbard-model}.
\newblock \emph{\bibinfo{journal}{Phys. Rev. B}} \textbf{\bibinfo{volume}{37}},
  \bibinfo{pages}{9753} (\bibinfo{year}{1988}).

\bibitem{Lindner10}
\bibinfo{author}{Lindner, N.}, \bibinfo{author}{Auerbach, A.} \&
  \bibinfo{author}{Arovas, D.~P.}
\newblock \bibinfo{title}{Vortex dynamics and hall conductivity of hard core
  bosons}.
\newblock \emph{\bibinfo{journal}{Phys. Rev. B}} \textbf{\bibinfo{volume}{82}},
  \bibinfo{pages}{134510} (\bibinfo{year}{2010}).

\bibitem{Avron85}
\bibinfo{author}{Avron, J.~E.} \& \bibinfo{author}{Seiler, R.}
\newblock \bibinfo{title}{Quantization of the hall conductance for general,
  multiparticle schr{\"o}dinger hamiltonians}.
\newblock \emph{\bibinfo{journal}{Phys. Rev. Lett.}}
  \textbf{\bibinfo{volume}{54}}, \bibinfo{pages}{259} (\bibinfo{year}{1985}).

\bibitem{Lindner09}
\bibinfo{author}{Lindner, N.~H.}, \bibinfo{author}{Auerbach, A.} \&
  \bibinfo{author}{Arovas, D.~P.}
\newblock \bibinfo{title}{Vortex quantum dynamics of two dimensional lattice
  bosons}.
\newblock \emph{\bibinfo{journal}{Physical Review Letters}}
  \textbf{\bibinfo{volume}{102}}, \bibinfo{pages}{070403}
  (\bibinfo{year}{2009}).

\bibitem{Simon83}
\bibinfo{author}{Simon, B.}
\newblock \bibinfo{title}{Holonomy, the quantum adiabatic theorem, and berry's
  phase}.
\newblock \emph{\bibinfo{journal}{Physical Review Letters}}
  \textbf{\bibinfo{volume}{51}}, \bibinfo{pages}{2167} (\bibinfo{year}{1983}).

\bibitem{Berry84}
\bibinfo{author}{Berry, M.~V.}
\newblock \bibinfo{title}{Quantal phase factors accompanying adiabatic
  changes}.
\newblock \emph{\bibinfo{journal}{Proc. R. Soc. Lond. A}}
  \textbf{\bibinfo{volume}{392}}, \bibinfo{pages}{45} (\bibinfo{year}{1984}).

\bibitem{Albuquerque07}
\bibinfo{author}{Albuquerque, A.} \emph{et~al.}
\newblock \bibinfo{title}{The alps project release 1.3: Open-source software
  for strongly correlated systems}.
\newblock \emph{\bibinfo{journal}{J. of Magn. Magn. Materials}}
  \textbf{\bibinfo{volume}{310}}, \bibinfo{pages}{1187} (\bibinfo{year}{2007}).

\bibitem{Fukui05}
\bibinfo{author}{Fukui, T.}, \bibinfo{author}{Hatsugai, Y.} \&
  \bibinfo{author}{Suzuki, H.}
\newblock \bibinfo{title}{Chern numbers in discretized brillouin zone:
  Efficient method of computing (spin) hall conductances}.
\newblock \emph{\bibinfo{journal}{J. Phys. Soc. Jpn.}}
  \textbf{\bibinfo{volume}{74}}, \bibinfo{pages}{1674} (\bibinfo{year}{2005}).

\bibitem{Huber08}
\bibinfo{author}{Huber, S.~D.}, \bibinfo{author}{Theiler, B.},
  \bibinfo{author}{Altman, E.} \& \bibinfo{author}{Blatter, G.}
\newblock \bibinfo{title}{Amplitude mode in the quantum phase model}.
\newblock \emph{\bibinfo{journal}{Phys. Rev. Lett.}}
  \textbf{\bibinfo{volume}{100}}, \bibinfo{pages}{050404}
  (\bibinfo{year}{2008}).

\bibitem{Simanek80}
\bibinfo{author}{Simanek, E.}
\newblock \bibinfo{title}{Instability of granular superconductivity}.
\newblock \emph{\bibinfo{journal}{Phys. Rev. B}} \textbf{\bibinfo{volume}{22}},
  \bibinfo{pages}{459} (\bibinfo{year}{1980}).

\bibitem{Sonin97}
\bibinfo{author}{Sonin, E.~B.}
\newblock \bibinfo{title}{Magnus force in superfluids and superconductors}.
\newblock \emph{\bibinfo{journal}{Phys. Rev. B}} \textbf{\bibinfo{volume}{55}},
  \bibinfo{pages}{485} (\bibinfo{year}{1997}).

\bibitem{Huber07}
\bibinfo{author}{Huber, S.~D.}, \bibinfo{author}{Altman, E.},
  \bibinfo{author}{B{\"u}chler, H.~P.} \& \bibinfo{author}{Blatter, G.}
\newblock \bibinfo{title}{Dynamical properties of ultracold bosons in an
  optical lattice}.
\newblock \emph{\bibinfo{journal}{Phys. Rev. B}} \textbf{\bibinfo{volume}{75}},
  \bibinfo{pages}{085106} (\bibinfo{year}{2007}).

\bibitem{Bakr09}
\bibinfo{author}{Bakr, W.~S.}, \bibinfo{author}{Gillen, J.~I.},
  \bibinfo{author}{Peng, A.}, \bibinfo{author}{F{\"o}lling, S.} \&
  \bibinfo{author}{Greiner, M.}
\newblock \bibinfo{title}{A quantum gas microscope for detecting single atoms
  in a hubbard-regime optical lattice}.
\newblock \emph{\bibinfo{journal}{Nature}} \textbf{\bibinfo{volume}{462}},
  \bibinfo{pages}{74} (\bibinfo{year}{2009}).

\bibitem{Jin96}
\bibinfo{author}{Jin, D.~S.}, \bibinfo{author}{Ensher, J.~R.},
  \bibinfo{author}{Matthews, M.~R.}, \bibinfo{author}{Wieman, C.~E.} \&
  \bibinfo{author}{Cornell, E.~A.}
\newblock \bibinfo{title}{Collective excitations of a bose-einstein condensate
  in a dilute gas}.
\newblock \emph{\bibinfo{journal}{Phys. Rev. Lett.}}
  \textbf{\bibinfo{volume}{77}}, \bibinfo{pages}{420} (\bibinfo{year}{1996}).

\end{thebibliography}
\end{document}